\documentclass[utf8]{FrontiersinHarvard} 

\usepackage{url,hyperref,lineno,microtype,subcaption}
\usepackage[doublespacing]{setspace}
\usepackage{threeparttable}
\usepackage{journalabb}

\def\keyFont{\fontsize{8}{11}\helveticabold }
\def\firstAuthorLast{Nakanotani {et~al.}} 
\def\Authors{Masaru Nakanotani\,$^{1,*}$, Lingling Zhao\,$^{1}$ and Gary P. Zank\,$^{1}$}
\def\Address{$^{1}$Center for Space Plasma and Aeronomic Research (CSPAR) and Department of Space Science, University of Alabama in Huntsville, Huntsville, AL, United States}
\def\corrAuthor{Masaru Nakanotani}

\def\corrEmail{mn0052@uah.edu}

\begin{document}
\onecolumn
\firstpage{1}

\title[Phase coherence of SW turbulence]{Phase Coherence of Solar Wind Turbulence from the Sun to Earth} 

\author[\firstAuthorLast]{\Authors} 
\address{} 
\correspondance{} 

\extraAuth{}

\maketitle

\begin{abstract}
The transport of energetic particles in response to solar wind turbulence is important for space weather.
To understand charged particle transport, it is usually assumed that the phase of the turbulence is randomly distributed (the random phase approximation) in quasi-linear theory and simulations.
In this paper, we calculate the coherence index, $C_\phi$, of solar wind turbulence observed by the Helios 2 and Parker Solar Probe spacecraft using the surrogate data technique to check if the assumption is valid.
Here, values of $C_\phi=0$ and $1$ indicate that the phase coherence is random and correlated, respectively.  
We estimate that the coherence index at the resonant scale of energetic ions ($10$ MeV protons) is $0.1$ at $0.87$ and $0.65$ au, $0.18$ at  $0.29$ au, and $0.3$ ($0.35$) at $0.09$ au for super (sub)-Alfv\'enic intervals, respectively. 
Since the random phase approximation corresponds to $C_\phi=0$, this may indicate that the random phase approximation is not valid for the transport of energetic particles in the inner heliosphere, especially very close to the Sun ($\sim0.09$ au).

\tiny
 \keyFont{ \section{Keywords:} Phase coherence, Solar wind turbulence, Inner heliosphere, and Surrogate data technique} 
\end{abstract}

\section{Introduction}
The propagation of solar energetic particles (SEPs) is important in the context of space weather \citep{malandraki2018}.
The Sun can be a frequent source of energetic particles ($>$ Mev) due to solar flares, shock waves associated with a coronal mass ejection, or the combination of both \citep{ryan2000,li2005}.
Since SEPs are a high radiation risk for astronauts working in space and for future human-exploration missions in the solar system, mitigation of SEPs is necessary to protect human health \citep{chancellor2014}.
Therefore, it is valuable to understand how energetic particles travel from the Sun to Earth and especially how long it takes for them to arrive so that mitigating actions can be taken, whether to evacuate astronauts or prevent damage to instruments on spacecraft.

The trajectory followed by energetic particles is a random walk thanks to scattering by solar wind turbulence \citep{vandenberg2020}.
According to \citet{zank1998, zhao2018}, the parallel mean free path of an energetic particle diffusing in response to the solar wind turbulence can be approximated as,
\begin{equation}
    \lambda_{||}\approx 6.2742\frac{B_0^{5/3}}{<\delta b^2>}\left(\frac{P}{c}\right)^{1/3}\lambda_{slab}^{2/3},
\end{equation}
where $B_0$ is the magnitude of the background magnetic field, $<\delta b^2>$ the turbulence magnetic energy, $P\equiv pc/Ze$ ($p$ momentum, $c$ the speed of light, and $Ze$ particle charge) the particle rigidity, and $\lambda_{slab}$ the correlation length of slab (or Alv\'enic) turbulence. 
Considering $10$ MeV protons as energetic particles, the rigidity becomes $P=137$ MV.
Typical solar wind turbulence parameters are $<\delta b^2>\sim100$ nT$^2$ and $\lambda_{slab}\sim0.5\times10^6$ km at $0.5$ au ($1$ au is the distance from the sun to Earth) \citep{adhikari2020}.
Using these parameters and $B_0\sim10$ nT, we obtain $\lambda_{||}\sim0.01$ au.
Since this is much smaller than $1$ au, the diffusion and transport processes of energetic particles are greatly affected by solar wind turbulence.
Furthermore, a perpendicular diffusion of energetic particles is also an important process \citep{zank2004, shalchi2010}.
The combination of parallel and perpendicular diffusion processes due to the turbulence makes estimates of the energetic particle arrival times complicated.

Although there are several studies of modeling and simulations to understand the diffusion and transport of solar energetic particles in turbulent magnetic fields, they usually assume that the phase of the turbulence is randomly correlated.
The quasi-linear diffusion theory is built on the assumption of the random phase approximation \citep{sagdeev1969}.
\citet{giacalone1994, otsuka2009, tautz2010, guo2014, moradi2022} investigate particle diffusion using test particle simulations combined with synthetic turbulence, which is described as a superposition of sine waves with random phases. 
The random phase approximation is useful and easily implemented in models and simulations.
However, the phase coherence of turbulence in the solar wind has not been addressed well from the perspective of observational data, especially close to the Sun.
While the intermittency of solar wind turbulence can be related to phase coherence \citep[and reference therein]{matthaeus2015}, which has been discussed frequently, in this paper, we directly measure the phase coherence of turbulence and focus on the link to particle diffusion and transport instead of intermittency.

Since it is possible that the phase coherence of turbulence modifies the motion of charged particles, it is valuable to consider whether the random phase approximation is reasonable in solar wind turbulence.
A good example of coherent turbulence (or structures) is short large amplitude magnetic structures (SLAMS) \citep{schwartz1991, schwartz1992, scholer1993}, which commonly form upstream of parallel shock waves due to the nonlinear evolution of ultra-low frequency waves excited by reflected ions.
These structures are quite coherent, and the amplitude is comparable to the background magnetic field \citep{koga2003}.
Some studies show that particle diffusion in coherent structures differs from the quasi-linear theory \citep{kirk1996, kuramitsu2000, hada2003b, laitinen2012}.
Note that \citet{kis2013} observed efficient ion acceleration in SLAMS at Earth's bow shock.
If solar wind turbulence is coherent, this can modify the diffusion and transport processes of energetic particles and the estimates of energetic particle arrival time at Earth can be different from estimates based on quasi-linear theory.
This motivates us to investigate the phase coherence of solar wind turbulence in the inner heliosphere.

In this paper, we investigate the phase coherence of solar wind turbulence from $0.09$ to $0.87$ au using Parker Solar Probe and Helios 2 observations.
These observations provide us with a unique opportunity to investigate solar wind turbulence in the inner heliosphere.
We use the surrogate data method to calculate the phase coherence in this study, which is explained in detail below.
Using the calculated phase coherence, we estimate phase coherence at a scale that is resonant with energetic particles and evaluate whether the phase random approximation is appropriate in the inner heliosphere.

\section{Method: Surrogate Data Technique}
The surrogate data technique \citep{hada2003a,koga2003} was developed originally to investigate the phase coherence of waves excited in the foreshock region of Earth's bow shock.
Suppose that $B_O(t)$ is an original time series obtained by a spacecraft. We can describe the method as follows: (I) decompose $B_O$ into the spectral amplitude $|\hat{B}_O(\omega)|$ and phase distribution $\phi(\omega)=\tan^{-1}\left[{\rm Im}(\hat{B}_O(\omega))/{\rm Re}(\hat{B}_O(\omega))\right]$, where $\hat{B}_O(\omega)$ is the Fourier transformation of $B_O(t)$. Here, we use a cosine cube-tapered rectangle function so that $2\%$ of the data time interval is tapered off at each end to minimize the effects of the boundary \citep{sahraoui2008}.
(II) We randomly shuffle the original phase distribution and perform an inverse Fourier transformation of the original spectral amplitude with the shuffled phase distribution.
We call this derived time series, $B_R(t)$, which is a phase-random surrogate.
(III) We repeat the same process as the previous one, but now use a correlated phase distribution ($\phi(\omega)=0$).
We call this, $B_C(t)$, a phase-correlated surrogate.
(IV) Once we obtain $B_O(t)$, $B_R(t)$, and $B_C(t)$, we calculate the first order structure function,
\begin{equation}
    L_j(\tau)=\sum_t|B_j(t+\tau)-B_j(t)|,
\end{equation}
where $j\in[O,R,C]$ and $\tau$ is the lag time.
(V) Finally, we evaluate the degree of phase coherence using the following equation \citep{sahraoui2008},
\begin{equation}
    C_\phi(\tau)=\frac{|L_O(\tau)-L_R(\tau)|}{|L_O(\tau)-L_R(\tau)|+|L_O(\tau)-L_C(\tau)|}.
	\label{eq:cphi}
\end{equation}
The phase coherence index, $C_\phi$, is an indication that when $C_\phi$ is close to $0$ ($1$), the phase coherence of the original data is random (correlated).
Note that Eq. \ref{eq:cphi} and the definition of $C_\phi$ in \cite{hada2003a} are equivalent when $L_R>L_O>L_C$.
It is useful to mention that it is not easy to determine whether the coherence is spatial or temporal since we use single-spacecraft data in this paper, and this can only be clarified using multi-point measurements, which are not manageable at this time in the inner heliosphere.

The original data $B_O(t)$ is calculated from the observed magnetic fields ${\bf B}(t)=(B_x, B_y, B_z)$ so that $B_O(t)$ corresponds to the largest relative fluctuation amplitude \citep{dudokdewit1996}.
We first compute ${\tilde B}(t,\varphi,\theta)={\bf B}(t)\cdot{\bf p}(\varphi,\theta)$ where ${\bf p}(\varphi,\theta)=(\cos\varphi\cos\theta, \sin\varphi\cos\theta, \sin\theta)$ is a unit vector in  spherical coordinates.
Here, the azimuthal angle $\varphi$ and elevation angle $\theta$ have a range of $-\pi\le\varphi\le\pi$ and $-\pi/2\le\theta\le\pi/2$, respectively.
When ${\tilde B}(t,\varphi,\theta)$ is projected onto the $\varphi-\theta$ plane by taking the time average, $<{\tilde B}(\varphi,\theta)>$, this gives us the distribution of the mean magnetic field.
The average $<{\tilde B}(\varphi,\theta)>=0$ corresponds to the direction ${\bf p}(\varphi,\theta)$ perpendicular to the mean magnetic field.
The projection of the normalized standard deviation can be calculated as,
\begin{equation}
    \sigma=\frac{\left<\left({\tilde B}-<{\tilde B}>\right)^2\right>^{1/2}}{<|{\bf B}|>}.
\end{equation}
Then, we find a direction $(\varphi_{\rm m}, \theta_{\rm m})$ that corresponds to the maximum value of $\sigma$ in the projection.
We use this direction to set $B_O(t)={\bf B}(t)\cdot{\bf p}(\varphi_m,\theta_m)$ for the following analysis unless otherwise stated, and this means that we calculate the coherence index for fluctuations corresponding to the largest fluctuation amplitude.
Note that since the choice of $B_O$ is arbitrary, for instance, it is also possible to set $B_O(t)$ as $|{\bf B}|-<|{\bf B}|>$.

In this paper, we choose well-studied turbulence cases in the inner heliosphere from Helios 2 and Parker Solar Probe observations.
Table 1 shows three Helios 2 observations at $0.87$, $0.65$, and $0.29$ au, respectively \citep{marsch1990}.
Magnetic field data were obtained by a flux-gate magnetometer (The Institute for Geophysics and Meteorology of the Technical University of Braunschweig magnetometer) \citep{musmann1977}.
The cadence of the magnetic field observations is $6$ s.
Each case is in the fast solar wind ($U>450$ kms$^{-1}$).
Observations shown in Table 2 were made by the Parker Solar Probe (PSP) spacecraft on 2021 April 28 during Encounter 8 \citep{zhao2022}.
The distance was at $0.09$ au.
Magnetic field data were obtained by the FIELDS Fluxgate Magnetometer instrument \citep{bale2016}. 
Here, we picked two cases of super- ($U>v_A$) and sub- ($U<v_A$) Alfv\'enic solar wind since the properties of solar wind turbulence are expected to be different across the Alfv\'en critical surface ($U=v_A$) \citep{kasper2021,zank2022}.
During this period, PSP observed several sub-Alfv\'enic flows \citep{kasper2021}.
We resampled the data down to a $1$ second resolution.
Note that we linearly interpolate missing data.

\begin{threeparttable}[htbp]
	\renewcommand{\arraystretch}{1.5}
	\caption{Helios 2 observations \citep{marsch1990}}
	\begin{tabular}{c c c c c c}\hline\hline
		Start Time [UT] & End Time [UT] & $R$ [au] & $U$ [km/s] & $v_A$ [km/s] & $B_0$ [nT]\\ \hline
		1976-02-19 00:00 & 1976-02-20 22:04 & 0.87 & 632 & 59  & 4.9\\ 
		1976-03-16 00:00 & 1976-03-17 00:00 & 0.65 & 621 & 79  & 8.36\\ 
		1976-04-15 00:00 & 1976-04-16 21:00 & 0.29 & 708 & 136 & 32.8\\ 
		\hline
	\end{tabular}
    \begin{tablenotes}
        \small
        \item \textbf{Note.} Values are adopted from \cite{marsch1990}. $R$ is the distance from the sun, $U$ the solar wind speed, $v_A$ the Alfv\'en speed, and $B_0$ the magnitude of the mean magnetic field.
    \end{tablenotes}
	\label{table:T1}
\end{threeparttable}

\begin{threeparttable}[htbp]
	\renewcommand{\arraystretch}{1.5}
	\caption{Parker Solar Probe observations \citep{zhao2022}}
	\begin{tabular}{c c c c c c c c}\hline\hline
		Start Time [UT] & End Time [UT] & $R$ [au] & $U$ [kms$^{-1}$] & $v_A$ [kms$^{-1}$] &$B_0$ [nT] & $\Psi$ [$^\circ$] & Interval \\ \hline
		2021-04-28 02:00 & 2021-04-28 07:00 & 0.09 & 345 & 257 & 234.9& 18 & super\\ 
		2021-04-28 09:33 & 2021-04-28 14:42 & 0.09 & 320 & 366 & 311.7& 15 & sub\\ 
		\hline
	\end{tabular}
    \begin{tablenotes}
        \small
        \item \textbf{Note.} Values are adopted from \cite{zhao2022}. $R$ is the distance from the sun, $U$ the solar wind speed, $v_A$ the Alfv\'en speed, $B_0$ the magnitude of the mean magnetic field, and $\Psi$ the angle between the mean magnetic field and solar wind speed.
    \end{tablenotes}
	\label{table:T2}
\end{threeparttable}

\section{Result}
\begin{figure}[h!]
\begin{center}
\includegraphics[width=15cm]{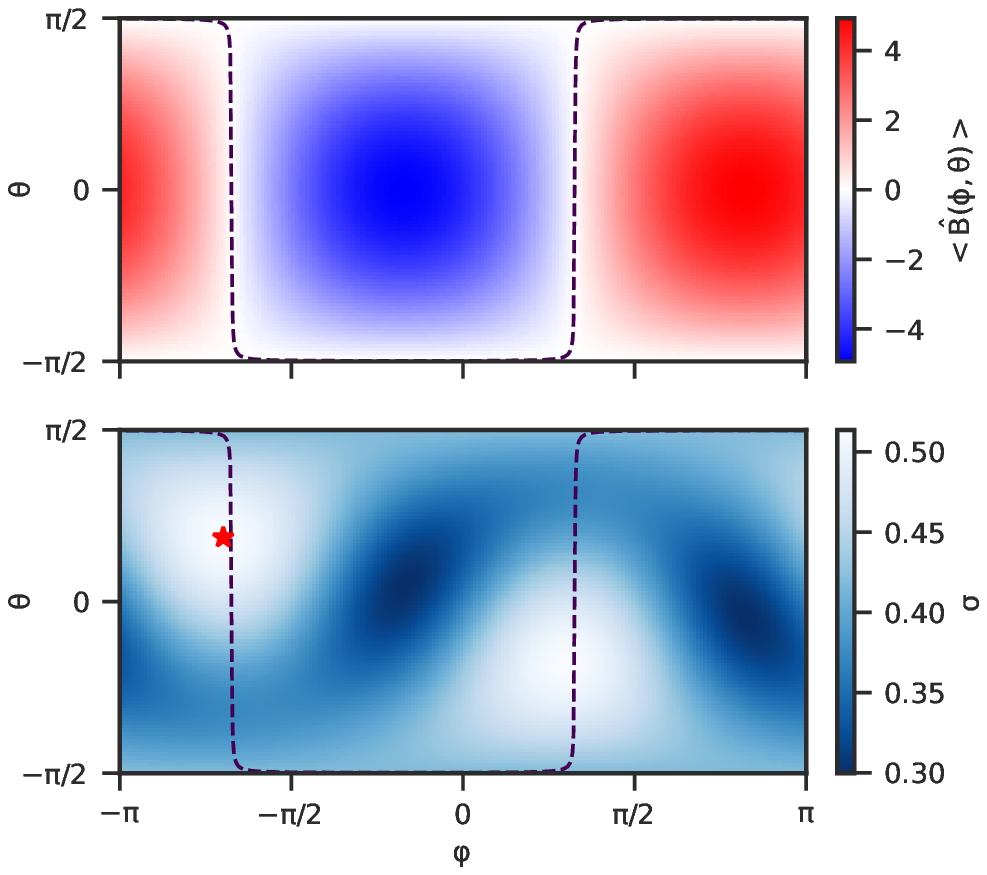}
\end{center}
\caption{Top panel: projection of the time average of $\hat{B}(t,\varphi,\theta)$ onto the $\varphi$-$\theta$ plane, and bottom panel: projection of the normalized standard deviation, $\sigma$ (see the text for the definition of $\sigma$). The dashed line indicates the direction perpendicular to the mean magnetic field. The star symbol corresponds to the maximum $\sigma$ value.}
\label{fig:prjctn}
\end{figure}

Fig. \ref{fig:prjctn} shows an example of a $(\varphi, \theta)$ projection at $0.87$ au.
The top panel is the mean magnetic field $<{\hat B}(\varphi, \theta)>$, and the dashed line corresponds to directions perpendicular to the mean magnetic field. 
The bottom panel displays the normalized standard deviation $\sigma$.
A large value of $\sigma$ means that the amplitude of fluctuations is large.
The star symbol in the bottom panel denotes the direction where the fluctuation amplitude is the maximum.
We can see that the maximum direction coincidently corresponds to a direction perpendicular to the mean magnetic field.
For other cases, not shown here, the maximum direction is also nearly perpendicular to the mean magnetic field.
Note that we also calculated the coherence index for other directions perpendicular to the mean magnetic field and found that the maximum direction tends to give the largest coherence index among other perpendicular directions (not shown here).

\begin{figure}[h!]
\begin{center}
\includegraphics[width=15cm]{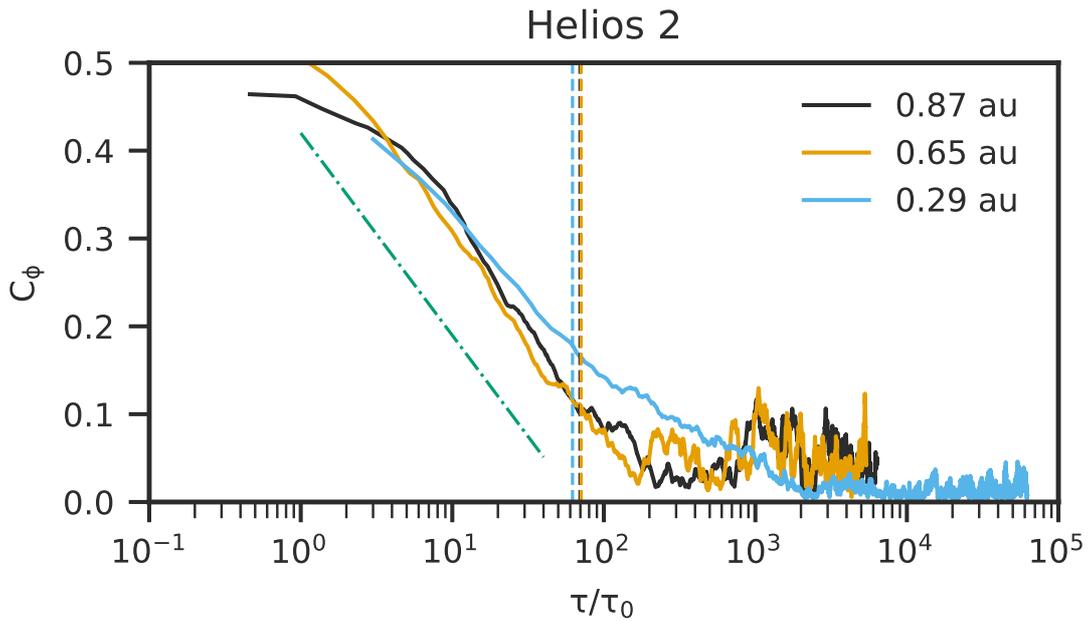}
\end{center}
\caption{Coherence index, $C_\phi$, calculated from Helios 2 observations (solid line). The dashed line corresponds to the resonant timescale of $10$ MeV protons. Black, orange, and blue colors correspond to distances at $0.87$, $0.65$, and $0.29$ au, respectively. The green dashed-dotted line is proportional to $\log\left[(\tau/\tau_0)^{0.1}\right]$. Here, $\tau_0$ is the local proton gyro period calculated at each distance.}
\label{fig:cphi_h2}
\end{figure}

The coherence index $C_\phi$ at $0.87$ au is shown as the black line in Fig. \ref{fig:cphi_h2}.
The index $C_\phi$ is moderately correlated ($C_\phi\sim0.45$) when the timescale of fluctuations is comparable to the local proton gyro period ($\tau\sim\tau_0$), and it gradually decreases as the fluctuation timescale increases and becomes randomly correlated ($C_\phi\sim0.05$) after $\tau/\tau_0=2\times 10^2$.
Here, the local proton gyro period $\tau_0$ is defined as $\tau_0=2\pi/\Omega_{cp}$ where $\Omega_{cp}$ is the proton cyclotron frequency.
Note that the decrease of $C_\phi$ in the range of $0.6<\tau/\tau_0<40$ can be fitted by $\log(\tau/\tau_0)^{-0.1}$.

\begin{figure}[h!]
\begin{center}
\includegraphics[width=15cm]{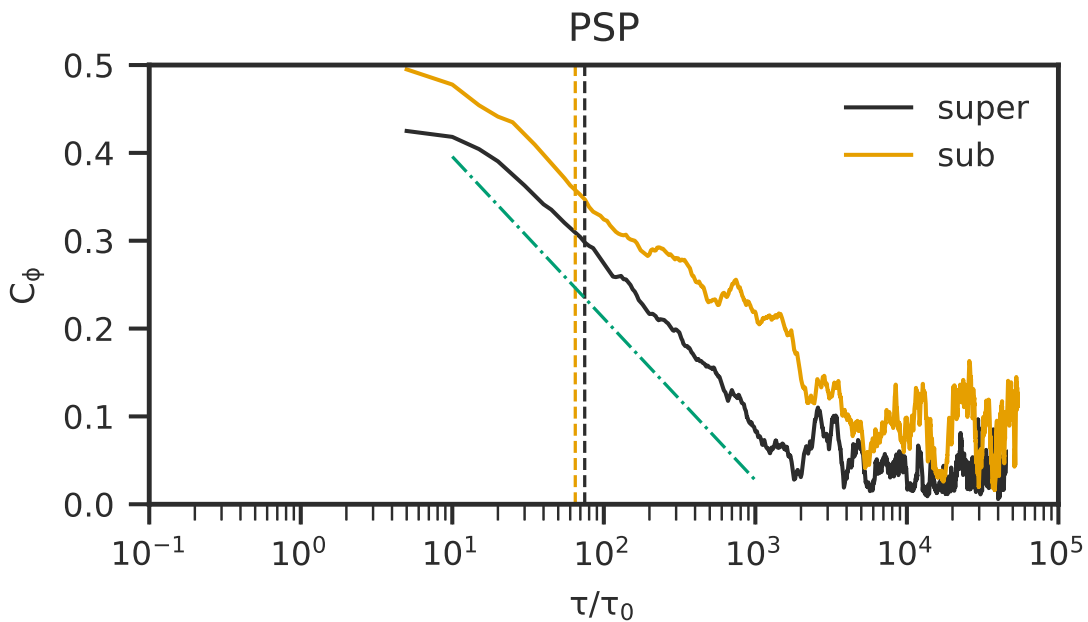}
\end{center}
\caption{Coherence index, $C_\phi$, calculated from PSP observations (solid line). The dashed line corresponds to the resonant timescale of $10$ MeV protons. Black and orange colors correspond to the super- and sub-Alfv\'enic intervals, respectively. The  green dashed-dotted line is proportional to $\log\left[(\tau/\tau_0)^{0.08}\right]$.Here, $\tau_0$ is the local proton gyro period calculated at $0.09$ au.}
\label{fig:cphi_ps}
\end{figure}

The profile of the coherence index is similar at smaller distances.
The orange and blue lines in Fig. \ref{fig:cphi_h2} show the coherence index at $0.65$ and $0.29$ au, respectively, and the black and orange lines in Fig. \ref{fig:cphi_ps} correspond to super- and sub-Alfv\'enic solar wind at $0.09$ au.
At $0.65$ au, the coherence index is $0.5$ when $\tau/\tau_0\sim1$, then gradually decreases until $\tau/\tau_0\sim2\times10^2$.
On the other hand, the coherence indices at $0.29$ and $0.09$ au converge after $\tau/\tau_0\sim10^3$.
The convergent value of $C_\phi$ at $0.29$ au is almost $0$, and that of $C_\phi$ at $0.09$ au is around $0.05$ and $0.1$ for super- and sub-Alfv\'enic solar wind, respectively.
Overall, the coherence index of the solar wind turbulence is moderately correlated when $\tau$ is comparable to $\tau_0$ and becomes random for large-scale magnetic fluctuations in the inner heliosphere.

It seems that the coherence index for small-scale fluctuations becomes larger when closer to the sun. 
From Fig. \ref{fig:cphi_h2}, the coherence index at $0.29$ au decreases more gradually than that at $0.87$ and $0.65$ au over the range of $1<\tau/\tau_0<10^3$.
The coherence index at $0.09$ au for the super-Alfv\'enic case is $0.4$ and $0.28$ at $\tau/\tau_0=10$ and $100$, respectively.
These values are larger than those found in Helios 2 observations.
This may indicate that fluctuations are more correlated closer to the sun and turn  more random as they propagate further in the solar wind.


The coherence indices at $0.09$ au for the super- and sub-Alfv\'enic intervals are noticeably different.
The black and orange lines in Fig. \ref{fig:cphi_ps} correspond to the super- and sub-Alfv\'enic intervals, respectively.
Although the overall profile is similar, we can see that the sub-Alfv\'enic $C_\phi$ is larger than the super-Alfv\'enic $C_\phi$ over all the fluctuation timescale.
This means that magnetic fluctuations are more correlated in the sub-Alfv\'enic region than the super-Alfv\'enic region.
Note that the decrease of $C_\phi$ in the range of $5<\tau/\tau_0<10^3$ for both cases can be fitted by $\log(\tau/\tau_0)^{-0.08}$.

The timescale for Alfv\'en waves resonant with energetic particles, $\tau_{res}$, can be approximated as $\tau_{res}/\tau_0=(c_p/U)(\Omega_{cp}/\Omega_{ci})$ and $(c_p/|U\cos\Psi+v_A|)(\Omega_{cp}/\Omega_{ci})$ for the Helios 2 and PSP observations, respectively.
Here, we assume that the observed fluctuations in the solar wind turbulence are dominated by Alfv\'en waves \citep{zank2022}.
We use Taylor's hypothesis for the Helios 2 observations, $\omega_{s/c}\sim kU$, since the solar wind speed is much higher than the Alfv\'en speed, $U\gg v_A$. 
Here, $\omega_{s/c}$ is the observed frequency in the spacecraft frame, and $k$ the wavenumber.
The resonant scale of energetic particles with Alfv\'en waves can be approximated as $\lambda_{res}\sim 2\pi c_p/\Omega_{ci}$ \citep{isenberg2005} where $c_p$ is the characteristic speed of energetic particles and $\Omega_{ci}$ the ion cyclotron frequency, and then the corresponding wavenumber is $k_{res}=2\pi/\lambda_{res}$.
Since the resonant frequency can be calculated as $\omega_{res}\sim k_{res}U$, the normalized resonant timescale is $\tau_{res}/\tau_0=(c_p/U)(\Omega_{cp}/\Omega_{ci})$.
For the PSP observations, we use a modified Taylor's hypothesis \citep{zank2022, zhao2022}, which describes $\omega_{s/c}\sim|U\cos\Psi\pm v_A|k$ where $\Psi$ is the angle between the mean magnetic field and the solar wind speed, since the solar wind speed is comparable to the Alfv\'en speed.
Here, $\pm$ corresponds to forward- and backward-propagating waves.
Using the same argument for $k_{res}$, we can write the resonant timescale as $\tau_{res}/\tau_0=(c_p/|U\cos\Psi+v_A|)(\Omega_{cp}/\Omega_{ci})$.
Here, we only consider forward-propagating waves since the forward waves are more abundant than the backward waves \citep{zank2022, zhao2022}.

The coherence index corresponding to the resonant timescale of $10$ MeV protons is at most $0.35$ in the inner heliosphere.
In the case of $10$ MeV protons, $\Omega_{ci}=\Omega_{cp}$ and $c_p\sim4.4\times10^4$ km/s.
The dashed lines in Fig. \ref{fig:cphi_h2} and \ref{fig:cphi_ps} correspond to the resonant timescale at each distance.
At the distance $R=0.87$ and $0.65$ au, the resonant timescales are almost the same, $\tau_{res}/\tau_0\sim70$, and the corresponding coherence index is $C_\phi\sim0.1$.
The observation of Helios 2 at $R=0.29$ au shows the resonant timescale, $\tau_{res}/\tau_0\sim62$, and the resonant coherence index is larger, $C_\phi=0.18$.
For the super- and sub-Alfv\'enic cases at $0.09$ au (black and orange dashed line), $\tau_{res}/\tau_0\sim75$ and $65$, respectively.
The corresponding $C_\phi$ are $0.3$ and $0.35$ for the super- and sub-Alfv\'enic intervals.
The smallest coherence index for $10$ MeV protons is $C_\phi\sim0.1$ at $0.87$ and $0.65$ au, and the largest coherence index is $C_\phi\sim0.35$ from the sub-Alfv\'enic case. 
This means that energetic particles interact with more correlated waves closer to the sun.

\begin{figure}[h!]
	\begin{center}
	\includegraphics[width=15cm]{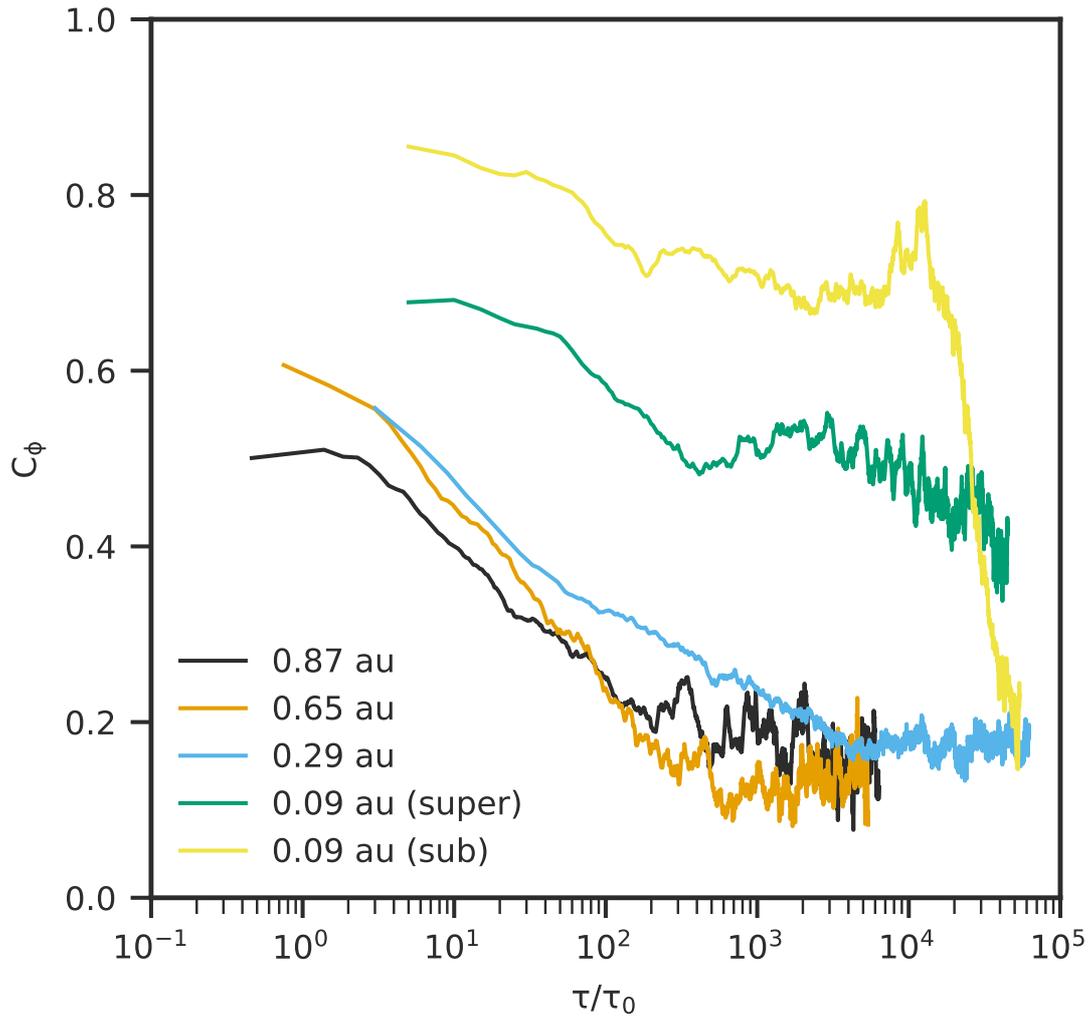}
	\end{center}
	\caption{Coherence index, $C_\phi$, calculated for fluctuations parallel to the mean magnetic field at each distance using Helios 2 and PSP observations.  Here, $\tau_0$ is the local proton gyro period calculated at each distance.}
	\label{fig:cphi_para}
\end{figure}

Phase coherence of fluctuations parallel to the mean magnetic field is found to be strong in the inner heliosphere.
While the above results are based on fluctuations perpendicular to the mean magnetic field, as suggested by the referee, we plot the coherence index of parallel fluctuations at each distance in Fig. \ref{fig:cphi_para}.
At $0.09$ au, the parallel coherence index in the sub-Alfv\'enic region is higher than $0.7$ over the entire range except for $\tau/\tau_0>2\times10^4$, indicating that the flucutuations are highly correlated, whereas in the super-Alfv\'enic region, that is decreased, but still shows a strong coherence ($0.4<C_\phi<0.7$ for $10<\tau/\tau_0<5\times10^4$).
The parallel coherence indices at $0.87$, $0.65$, and $0.29$ au are similar to the perpendicular coherence indices shown above, but slightly higher than them, and converge to around $0.2$.
It is also noticeable that the parallel coherence index increases with reduced distance to the sun.
It is not clear the reason of the strong coherence for parallel fluctuations in the inner heliosphere, this needs a further investigation.

\section{Summary and Discussion}
We have calculated the coherence of the solar wind turbulence in the inner heliosphere using the surrogate data technique.
Well-known turbulence studies of Helios 2 and PSP observations \citep{marsch1990, zhao2022} are used in this paper.
The coherence index $C_\phi$ calculated by the surrogate data technique shows that $C_\phi$ is $\sim0.45$ when the timescale of fluctuations, $\tau$, is comparable to the local gyro period and converges to less than $0.1$ as $\tau$ increases at $0.87$ au.
We found that the coherence index tends to be larger as the distance to the Sun decreases.
There is an evident difference in the coherence index between the super- and sub-Alfv\'en intervals, showing that the sub-Alfv\'enic $C_\phi$ value is larger than the super-Alfv\'enic $C_\phi$ value over the entire range of $\tau$.

Compared to the coherence index at the resonant timescale of energetic ions ($>10$ MeV), the random phase approximation can be regarded as valid in the inner heliosphere.
We have assumed that the observed fluctuations are dominated by Alfv\'en waves, and computed the resonant scale of $10$ MeV protons at each distance and the corresponding coherence index.
We found that the coherence index is $0.1$ at $0.87$ and $0.65$ au, $0.18$ at $0.29$ au, $0.3$ and $0.35$ at $0.09$ au for the super- and sub-Alfv\'enic intervals, respectively.
Overall, the coherence index for $10$ MeV protons is still finite in the inner heliosphere.
This is consistent with previous observations at Earth's bow shock \citep{koga2007, koga2008}.
Since the random phase approximation used in the quasi-linear theory and simulations corresponds to $C_\phi=0$, this may suggest that the random phase approximation is violated for the transport of energetic particles in the inner heliosphere, especially very close to the Sun ($\sim0.09$ au).
Note that more energetic particles ($>10$ MeV) interact with fluctuations less correlated since the resonant timescale becomes larger and the coherence index tends to be smaller for large $\tau$.
Since the coherence index at the resonant timescale is finite, it may require us to include the effects of phase coherence of turbulence in theories and simulations to fully understand the transport of energetic particles in finitely-correlated fluctuations.
Furthermore, since the coherence of parallel fluctuations is also finite and seemingly strong, we may need to consider finite coherence for the transport of energetic particles due to parallel fluctuations.
However, we have to keep in mind that the amplitude of parallel fluctuations is generally small compared to that of perpendicular fluctuations (see Fig. \ref{fig:prjctn} for instance).


Although we have assumed that the observed fluctuations are Alfv\'enic, we may need to consider the contribution of 2D modes to the transport of energetic particles.
It has been pointed out that solar wind turbulence contains 2D modes as well as Alfv\'en waves.
For 2D modes, the resonant scale for energetic particles can be considered as the Larmor radius, $\lambda_{res}\sim c_p/\Omega_{ci}$, and this yields a larger resonant timescale.
Therefore, this corresponds to a smaller coherence index.
Since the surrogate data technique used here cannot distinguish Alfv\'en waves and 2D modes, this needs further investigations.

Future work will examine more intervals of solar wind turbulence for better statistics and to strengthen our conclusion.
Since we focus only on fast solar wind in this paper, it would be interesting to investigate the case of slow solar wind.
Furthermore, it would be interest to investigate the coherence index in the outer heliosphere using Voyager spacecraft data.

\section*{Conflict of Interest Statement}

The authors declare that the research was conducted in the absence of any commercial or financial relationships that could be construed as a potential conflict of interest.

\section*{Author Contributions}
MN performed the analysis of the observational data and prepared the first draft. 
GZ and LZ contributed to the discussion of the method, results, and the preparation of the article.

\section*{Funding}
We acknowledge the partial support of an NSF EPSCoR RII
Track-1 Cooperative Agreement OIA-2148653, partial support
from a NASA Parker Solar Probe contract SV4-84017, partial
support from NASA awards 80NSSC20K1783 and 80NSSC23K0415, and partial
support from a NASA IMAP sub-award under NASA contract
80GSFC19C0027. The SWEAP Investigation and this study are
supported by the PSP mission under NASA contract
NNN06AA01C.

\section*{Acknowledgments}
The Parker Solar Probe was designed, built, and is now operated by the Johns Hopkins Applied Physics Laboratory as part of NASA’s Living with a Star (LWS) program (contract NNN06AA01C). Support from the LWS management and technical team has played a critical role in the success of the Parker Solar Probe mission.
We thank the NASA Parker Solar Probe FIELDS team led by S. D. Bale for use of data.
We are grateful to the referees for their thoughtful reports and valuable comments.


\section*{Data Availability Statement}
Helios 2 data are publicly available from NASA’s Space Physics Data Facility (SPDF) at https://spdf.gsfc.nasa.gov/pub/data/helios/helios2


\bibliographystyle{Frontiers-Harvard} 
\bibliography{test}


\section*{Figure captions}


\end{document}